\documentclass[12pt]{article}
\usepackage{mciteplus}
\usepackage{authblk}
\usepackage[T1]{fontenc}
\usepackage{lmodern}
\usepackage{lscape}
\usepackage[export]{adjustbox}
\usepackage{numprint}
\usepackage{multirow}
\usepackage{ifthen} 
\newboolean{pdflatex}
\setboolean{pdflatex}{true} 

\newboolean{articletitles}
\setboolean{articletitles}{true} 
\usepackage[bookmarksnumbered, colorlinks, plainpages]{hyperref}
\usepackage{amsmath, amsthm, amscd, amsfonts, amssymb, graphicx, color, booktabs}
\usepackage[square,sort,comma,numbers]{natbib}

\numberwithin{equation}{section}
\definecolor{email}{rgb}{0.00,0.00,0.84}
\begin{document}
\setcounter{page}{1}

\title{\large \bf 12th Workshop on the CKM Unitarity Triangle\\ Santiago de Compostela, 18-22 September 2023 \\ \vspace{0.3cm}
\LARGE Handling Correlated Systematic Uncertainties on the CKM Angle $\gamma$}

\author[1,2]{Alex Gilman}
\affil[1]{University of Oxford, Oxford, United Kingdom}
\affil[2]{on behalf of the LHCb Collaboration}
\maketitle

\begin{abstract}
The current precision on the CKM angle $\gamma$ is driven by averages of measurements from multiple final states and multiple experiments. As the next generation of experiments targets a total experimental precision on $\gamma$ less than one degree, systematic uncertainties must be well-controlled and correlated uncertainties between different final states and across experiments must be fully understood. These proceedings examine the systematic uncertainties that arise in direct measurements of $\gamma$ and discuss the potential of each class of systematic uncertainties to prevent sub-degree precision.
\end{abstract} \maketitle


\noindent The CKM angle $\gamma$ is a key parameter in testing the hypothesised unitarity of the CKM matrix and, in combination with measurements of $|V_{ub}|$, provides a test of unitarity with processes that proceed only through tree-level interactions. There is negligible theoretical uncertainty in extracting $\gamma$ from measured quantities, and thus the precision in its determination is entirely limited by statistical and systematic uncertainties from experiment. The precision on $\gamma$ from direct measurements is driven by measurements of interference between $b\to c\overline u  s$ and $b\to \overline c u  s$ transitions, with the decay $B^+\to D K^+$ providing leading precision, where $D$ represents a superposition of a $D^0$ and $\overline D^0$ states. Ultimate precision on $\gamma$ is achieved by examining multiple $D$ final states, such as $D\to K^+K^-$, $D \to K^\pm \pi^{\mp}(\pi^+\pi^-)$, and $D\to K_S^0\pi^+\pi^-$, some of which rely on inputs of measured $D$-decay hadronic parameters from dedicated charm experiments, namely CLEO-c and BESIII. Measurements of mixing in the $D^0-\overline D^0$ system are also sensitive to these same parameters and significant precision on some $D^0$ hadronic parameters is achieved in mixing measurements from LHCb. While measurements of $B^+\to D K^+$ provide leading precision on $\gamma$, competitive precision can be achieved by averaging measurements of various $B\to D h$-like decays to the same set of $D$ final states, which introduces correlated uncertainties as these measurements depend on the same set of $D$-decay hadronic parameter measurements to extract $\gamma$. 

\begin{figure} [hbt!]
\centering
\includegraphics[height=0.3\textwidth]{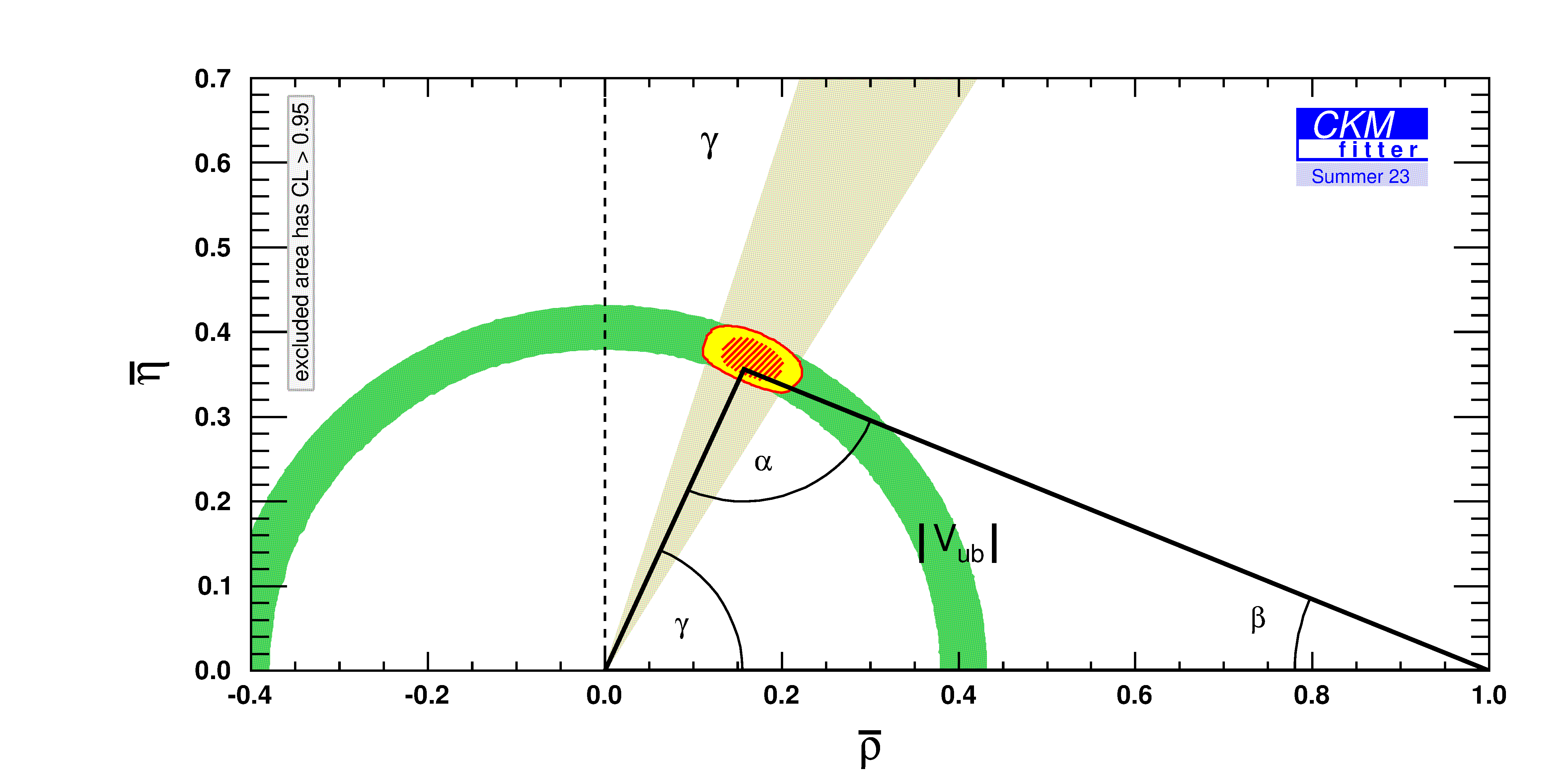}\includegraphics[height=0.3\textwidth]{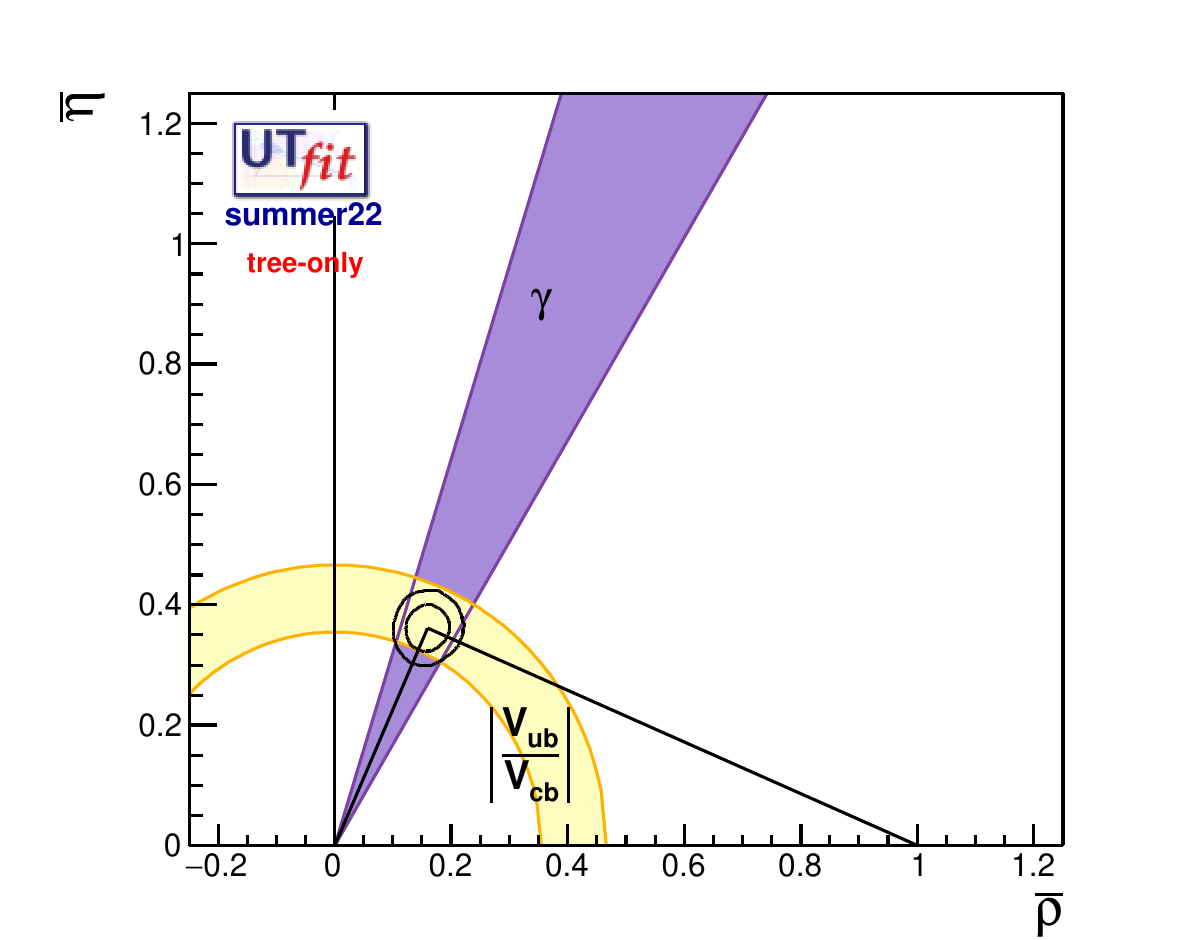}
 \caption{Current status of CKM-unitarity tests from averages of direct measurements of $\gamma$ and $|V_{ub}|$ from (left) the CKMFitter Collaboration~\cite{CKMFitter} and (right) the UTFit collaboration~\cite{UTFit}. }
\label{fig:Avgs}
\end{figure}

The current status of CKM unitarity from tree-level measurements by the CKMFitter collaboration~\cite{CKMFitter} and the UTFit collaboration~\cite{UTFit} is shown in Fig.~\ref{fig:Avgs}. An average performed by the LHCb collaboration in 2022 of all LHCb measurements of $\gamma$, which contribute most significantly to current precision in global averages, found $\gamma=(63.8^{+3.5}_{{-3.7}})^\circ$, with the contribution from systematic uncertainties totalling approximately $1.4^\circ$~\cite{LHCbCombo}. The CKMFitter collaboration predicts sub-degree level precision on $\gamma$ by the end of Run4 at the LHC~\cite{CKMFitterProjection}. In order to reach that goal, the total systematic uncertainty must be controlled to much less than one degree. 

\begin{figure} [hbt!]
\centering
\includegraphics[width=0.50\textwidth]{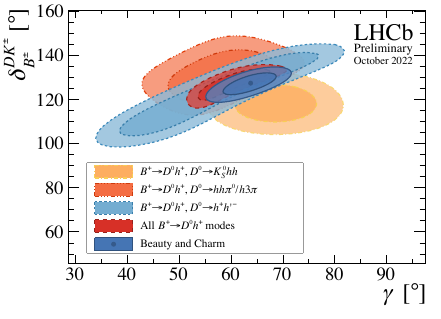}
 \caption{Constraints from LHCb measurements on the parameter space of $\gamma$ versus 
 $\delta_{B^\pm}^{DK^\pm}$, the strong phase of the $B^+\to D K^+$ decay, by $D$ final state. }
\label{fig:LHCbCombo}
\end{figure}

The discussion on systematic uncertainties on direct measurements of $\gamma$ in these proceedings is structured around measurements of $B^+\to D K^+$ with the three sets of $D$ final states which provide leading precision: the two-body Gronau-London-Wyler (GLW) decays~\cite{GLW1,GLW2} ($D\to \pi^+\pi^-$ and $D\to K^+K^-$) and Atwood-Dunietz-Soni (ADS) decays~\cite{ADS}  $D\to K^\pm \pi^{\mp}$, the Bondar-Poluetkov-Giri-Grossman-Soffer-Zupan (BPGGSZ) decays~\cite{ggsz1,ggsz2,ggsz3,ggsz4}  $D\to K_S^0 \pi^+\pi^{-}$, and the binned analysis of the four-body ADS decays $D\to K^\pm\pi^{\mp}\pi^+\pi^-$. The relative contributions of each LHCb measurement to the LHCb average is shown in Fig.~\ref{fig:LHCbCombo}.

Measurements of the two-body ADS and GLW modes in $B^+\to DK^+$ decays were published by the LHCb collaboration in 2021~\cite{LHCbTwoBody} and a combined analysis of the same states with Belle and Belle II data was released on the arXiv in late 2023~\cite{BelleTwoBody}. A summary of the size of systematic uncertainties in the LHCb measurement is shown in Table~\ref{tab:HHUncertainties}. In both measurements, the systematic uncertainties are dominated by estimations of charmless $B$ decays to the same final state. However, both measurements employ data control samples to study these backgrounds and the assigned uncertainties are solely due to limited statistical precision from the size of the control samples and so are uncorrelated across the experiments. The LHCb measurement also assigns sizable systematic uncertainties to the estimation of misidentified $B_s^0$ and $\Lambda_b^0$ decays, which are not present in the Belle + Belle II data. One can expect these uncertainties to scale with the size of the collected LHCb data sample, pending further analysis of these backgrounds. The only dependence on external inputs for these measurements is from the hadronic parameters of $D\to K^{\mp}\pi^{\pm}$, which are included as free parameters in global averages. Correlations from the $D\to K^{\mp}\pi^{\pm}$ input parameters are then handled implicitly, and the incorporation of future measurements that constrain $D\to K^{\mp}\pi^{\pm}$ hadronic parameters should be relatively straightforward. 

\begin{table}[hbtp]
    \centering
        \caption{Summary of systematic uncertainties on CPV observables which drive precision on $\gamma$ from the LHCb measurement of $B^+\to D K^+$, $D\to h^+{h'}^- $. }
    \begin{tabular}{c|c}
    CPV Observable& $\frac{\text{Sys. Uncertainty}}{\text{Stat. Uncertainty}}$ $(\%)$\\\hline
         $A_K^{CP}$&  16\\
         $R^{CP}$ &   106 \\
         $R^{\pi K}_{K^-}$ & 57  \\
         $R^{\pi K}_{K^+}$ &   53
    \end{tabular}
    \label{tab:HHUncertainties}
\end{table}

Measurements of $B^+\to D K^+$ with $D\to K_S^0 \pi^+\pi^-$ have also been published by both the LHCb collaboration in 2020~\cite{LHCbGGSZ} and Belle and Belle II collaborations in 2022~\cite{BelleGGSZ}. A table summarising the uncertainties of the LHCb measurement is included in Table~\ref{table:GGSZUncertainties}. The systematic uncertainties on the Belle and Belle II measurement are of the same order, with roughly $3.5$ times larger statistical uncertainties.  Due to the different experimental setups, the $B^+$-related related systematic uncertainties between the LHCb measurement and the Belle and Belle II measurement are largely uncorrelated. The LHCb-related systematic uncertainty is not dominated by any single-source, but the largest sources of systematic uncertainty relate to the variation of efficiencies and detector resolution across the phase space and modelling partially-reconstructed backgrounds. The background-related uncertainties should scale with the size of data in a straightforward fashion, but the other uncertainties require more detailed study for significant reduction.

\begin{table}[hbtp]
    \centering
        \caption{Summary of uncertainties on CPV parameters from the LHCb measurement of $B^+\to D K^+$ with $D\to K_S^0 h^+h^-$~\cite{LHCbGGSZ}.}
    \begin{tabular}{c|cccc}
         Source & $\sigma(x_-^{DK})$ & $\sigma(xy_-^{DK})$ & $\sigma(x_+^{DK})$ & $\sigma(y_+^{DK})$ \\ \hline \hline
        Statistical & 0.96 & 1.14 & 0.98 &1.23\\\hline
        $D$ hadronic parameter inputs & 0.23 & 0.35 & 0.18 & 0.28 \\\hline
        Total $B^+$-related uncertainty& 0.20 & 0.25 &0.24 &0.26\\\hline
        Total systematic uncertainty & 0.31 &0.43 &0.30 & 0.38
    \end{tabular}
    \label{table:GGSZUncertainties}
\end{table}

As the table shows, the contributions from external hadronic parameter inputs and $B^+$-related related systematic uncertainties are of the same order. These analyses fix the hadronic parameter inputs in the fits, and so reinterpreting the results with updated measurements of the hadronic parameters is difficult. Additionally, as both analyses (and other analyses of $B\to Dh$-like decays with $D\to K_S^0\pi^+\pi^-$) employ the same set of hadronic parameter inputs, the uncertainties related to these inputs are expected to be correlated across $B\to Dh$-like measurements. However, due to sensitivity being driven by different regions of phase space in different $B\to Dh$ channels and different efficiency profiles, these correlations may be significantly less than unity. The 2023 LHCb publication of $B^0\to D K^{*}(892)^0$ with $D\to K_S^0 h^+ h^-$~\cite{B0GGSZ} reports the correlation of hadronic parameter uncertainties with the measurement of $B^+\to D K^{+}$, and indeed finds that the correlations deviate significantly from unity, as demonstrated in Table~\ref{table:GGSZCorrs}. Correlations between analyses that follow the same procedure to Ref.~\cite{LHCbGGSZ} can determine correlation coefficients of these uncertainties in a similar fashion pending the publication of the variations of the hadronic parameters employed in the study. An alternative publication strategy could allow for these correlations to be handled implicitly in global averages, namely publishing the yields of the $B\to DK$-like and $B^+\to D\pi^+$ in each $D\to K_S^0h^+h^-$ bin.

\begin{table}[h]
    \centering
    \caption{Correlation matrix of hadronic parameter systematic uncertainties on CPV observables from $B^0\to DK^{*0}(892)$~\cite{B0GGSZ} (with superscript $DK^{*0}$)  and \mbox{$B^+\to DK^{+}$}~\cite{LHCbGGSZ} (with superscript $DK$) with $D\to K_S^0 h^+h^-$.}
    \small
    \scalebox{0.9}{
    \begin{tabular}{l|rrrrrrrrrr}
        & $x_+^{DK^{*0}}$ & $x_-^{DK^{*0}}$ & $y_+^{DK^{*0}}$ & $y_-^{DK^{*0}}$ & $x_-^{DK}$ & $x_+^{DK}$ & $y_-^{DK}$ & $y_+^{DK}$ \\
        \hline
$x_+^{DK^{*0}}$ & {1.00} & {$-$0.14} & {0.34} & {$-$0.09} & {$-$0.29} & {$-$0.06} & {$-$0.11} & {$-$0.06} \\
$x_-^{DK^{*0}}$ & & {1.00} & {$-$0.04} & {0.17} & {$-$0.31} & {0.48} & {0.22} & {$-$0.49} \\
$y_+^{DK^{*0}}$ & & & {1.00} & {$-$0.04} & {0.35} & {0.03} & {0.12} & {0.27}  \\
$y_-^{DK^{*0}}$ & & & & {1.00} & {0.13} & {$-$0.15} & {0.22} & {$-$0.01}  \\
$x_-^{DK}$       & & & & & {1.00} & {$-$0.49} & {$-$0.05} & {0.32}  \\
$x_+^{DK}$       & & & & & & {1.00} & {0.06} & {0.06} \\
$y_-^{DK}$       & & & & & & & {1.00} & {$-$0.24}  \\
$y_+^{DK}$       & & & & & & & & {1.00}  \\
    \end{tabular} 
    }
    \label{table:GGSZCorrs}
\end{table}

In 2023, LHCb published an analysis of $B^{\pm}\to D K^{\pm}$, with $D\to K^\mp \pi^\pm \pi^+\pi^-$~\cite{LHCbK3Pi} based on a binning scheme of the $D\to K^\mp \pi^\pm \pi^+\pi^-$ phase space suggested in Ref.~\cite{K3PiBinning}. This analysis determined $\gamma=\left(54.8^{+6.0}_{-5.8}\,^{+0.6}_{-0.6}\,^{{+6.7}}_{{-4.7}}\right)^\circ$, where the first uncertainty is the statistical uncertainty of the determined number of $B^{\pm}\to D K^{\pm}$ in the LHCb sample, the second uncertainty is LHCb-related systematic uncertainty, and the third uncertainty is the propagated uncertainty on the input hadronic parameters from a BESIII measurement of $D\to K^\mp \pi^\pm \pi^+\pi^-$ decays~\cite{BESIIIK3Pi}. Due to limited BESIII statistics and statistical correlations from the interpretation of the BESIII data, the likelihood profile of the eight-dimensional $D\to K^\mp \pi^\pm \pi^+\pi^-$ hadronic parameter space (two parameters in each of the four bins) is highly non-trivial. Any other measurements with this channel would also be limited by the same uncertainty, and thus further precision on the inputs is required for significantly improved precision.

Some common sources of uncertainty are present across LHCb measurements of $B^{\pm}\to D K^{\pm}$ with different $D$ decay channels, the most notable source being the modelling of partially reconstructed $B^{\pm}\to D^{*0} K^{\pm}$ decays. However, the dominating sources of uncertainty from $B^+$ CPV measurements across $D$ decay channels are largely uncorrelated. In measurements of $B^+\to D K^+$, production and detection asymmetries are controlled by reference $B^+\to D \pi^+$ channel, and thus rely on independent statistical uncertainties across different $D$ decay modes. As such, the only appreciable source of correlated uncertainty to be accounted for in global averages arises from common inputs for $D$ hadronic parameters. New data collected by the BESIII will provide roughly seven times as many $D^0$ mesons for analysis compared to the previous world's largest dataset, and so will significantly reduce the uncertainties on these common input parameters, and uncertainties on the measured hadronic parameters are projected to scale with square root of the increased BESIII statistics.

As mentioned previously, the measurement of $B^+\to D K^+$ with $D\to K_S^0 h^+h^-$ events from Ref.~\cite{LHCbGGSZ} does not easily lend itself to reinterpretation with new measurements of the input hadronic parameters, as the parameters are fixed in the fits performed to $B^+$ data. LHCb plans to address this by publishing the determined signal yields in each bin of parameter space and the related correlation matrices, as done in the analysis of $B^+\to D K^+$ with $D\to K^+K^-\pi^+\pi^-$~\cite{KKPiPi}.

In summary, while measurements of $\gamma$ are still statistically limited, systematic uncertainties need to be reduced to achieve sub-degree precision. Systematic uncertainties from the measurements of CPV in beauty hadrons are largely uncorrelated, but significant correlations are introduced in the interpretation of these results due to shared inputs of $D$ hadronic parameters. New data from BESIII will provide the required measurements to not impede sub-degree precision.

\section*{Acknowledgments}
Financial support by the European Commission through the grant ONEDEGGAM, Grant number 758462 is gratefully acknowledged.


\vspace{-1ex}

\addcontentsline{toc}{section}{References}
\setlength{\bibsep}{0.0pt}
\bibliographystyle{LHCb}
\bibliography{mybib}


\end{document}